\documentclass[conference]{IEEEtran}

\usepackage[pdftex]{graphicx}
\usepackage{xcolor}
\usepackage[hidelinks]{hyperref}

\usepackage{balance}

\definecolor{Paired-1}{RGB}{31,120,180}
\definecolor{Paired-2}{RGB}{166,206,227}
\definecolor{Paired-3}{RGB}{51,160,44}
\definecolor{Paired-4}{RGB}{178,223,138}
\definecolor{Paired-5}{RGB}{227,26,28}
\definecolor{Paired-6}{RGB}{251,154,153}
\definecolor{Paired-7}{RGB}{255,127,0}
\definecolor{Paired-8}{RGB}{253,191,111}
\definecolor{Paired-9}{RGB}{106,61,154}
\definecolor{Paired-10}{RGB}{202,178,214}
\definecolor{Paired-11}{RGB}{177,89,40}
\definecolor{Paired-12}{RGB}{255,255,153}

\usepackage{amsmath,amssymb,amsfonts,stmaryrd}
\usepackage{bm}
\usepackage{dsfont}
\usepackage{numprint}  
\newcommand{\isep}{\mathrel{{.}\,{.}}\nobreak}

\usepackage{adjustbox}
\usepackage{booktabs}
\usepackage{multirow}

\usepackage[caption=false,font=footnotesize]{subfig}
\usepackage{tikz}
\usetikzlibrary{matrix, positioning, patterns, shapes, arrows}
\usepackage{pgfplots}
\usepgfplotslibrary{groupplots}
\usepgfplotslibrary{colorbrewer}

\usepackage{pgfplots}
\pgfplotsset{compat=newest}
\usepgfplotslibrary{groupplots}

\usepackage{tikz}
\usetikzlibrary{matrix, positioning, patterns, shapes, arrows}

\usepackage[draft]{fixme} 
\fxsetup{theme=color}

\usepackage{lipsum}

\usepackage[ruled, linesnumbered, vlined]{algorithm2e}

\title{High-Throughput VLSI Architecture for GRAND Markov Order}

\author{\IEEEauthorblockN{Syed Mohsin Abbas, Marwan Jalaleddine and Warren J. Gross}
\IEEEauthorblockA{Department of Electrical and Computer Engineering\\
McGill University,  Montr\'eal, Qu\'ebec, Canada\\
Emails: syed.abbas@mail.mcgill.ca, marwan.jalaleddine@mail.mcgill.ca, warren.gross@mcgill.ca}}

\begin{document}

\maketitle

\begin{abstract}
Guessing Random Additive Noise Decoding (GRAND) is a recently proposed Maximum Likelihood (ML) decoding technique. Irrespective of the structure of the error correcting code, GRAND tries to guess the noise that corrupted the codeword in order to decode any linear error-correcting block code. GRAND Markov Order (GRAND-MO) is a variant of GRAND that is useful to decode error correcting code transmitted over communication channels with memory which are vulnerable to burst noise. Usually, interleavers and de-interleavers are used in communication systems to mitigate the effects of channel memory. Interleaving and de-interleaving introduce undesirable latency, which increases with channel memory. To prevent this added latency penalty, GRAND-MO can be directly used on the hard demodulated channel signals. This work reports the first GRAND-MO hardware architecture which achieves an average throughput of up to $52$ Gbps and $64$ Gbps for a code length of $128$ and $79$ respectively. Compared to the GRANDAB, hard-input variant of GRAND, the proposed architecture achieves $3$ dB gain in decoding performance for a target FER of $10^{-5}$. Similarly, comparing the GRAND-MO decoder with a decoder tailored for a $(79,64)$ BCH code showed that the proposed architecture achieves 33$\%$ higher worst case throughput and $2$ dB gain in decoding performance.

\end{abstract}

 
\section{Introduction}
For 5G and beyond communication networks, ultra-reliable low-latency communication (URLLC) \cite{URLLC2} is a very promising addition to the pre-existing communication standards \cite{3GPP}. URLLC enables many applications such as augmented and virtual reality,  intelligent transportation systems (ITA) \cite{URLLC1}, internet of things (IoT) \cite{IoT1,IoT2}, machine to machine communication and many others \cite{URLLC3}. Realizing these applications requires short high-rate maximum likelihood performing codes to support the low latency and high reliability requirements of mission critical events. For such codes, GRAND has been developed as a maximum likelihood (ML) decoding algorithm \cite{Duffy19TIT}. GRAND attempts to guess the noise that corrupted the transmitted codeword rather than decoding the received vector by leveraging the structure of the underlying code. This makes GRAND a desirable code agnostic decoder as it can be used to decode any linear block code. GRAND relies on the generation of putative test error patterns that are successively applied to the received vector. The order in which these putative test error patterns are generated is the key difference between different variants of GRAND. There are hard-input variants of GRAND (GRANDAB) \cite{Duffy19TIT}  as well as soft-input variants (ORBGRAND \cite{duffy2020ordered}, SRGRAND \cite{SRGRAND}, SGRAND \cite{solomon2020soft}). 
GRAND Markov Order (GRAND-MO) \cite{GRANDMO} is a hard-input GRAND variant designed specifically for channels with memory which are susceptible to burst noise. Due to the effect of burst noise, channels with memory suffer from a significant degradation in decoding performance with typical channel code decoders, and this degradation increases with channel memory \cite{GRANDMO}. As a result, interleavers/deinterleavers are used to mitigate the effects of burst noise in order to reduce performance degradation. Interleavers and deinterleavers, on the other hand, introduce additional latency. In emerging applications such as URLLC \cite{URLLC2}-\cite{URLLC3}, where latency and reliability are critical, the delay imposed by interleavers/de-interleavers or the performance degradation caused by channel memory is unacceptable. GRAND Markov Order (GRAND-MO) \cite{GRANDMO} eliminates the need for interleavers/deinterleavers for channels with memory, allowing for effective and reliable communication in the presence of burst noise. GRAND-MO makes use of noise correlations and adapts its test error pattern generation to mitigate the effect of noise bursts. As a result, GRAND-MO outperforms traditional channel code decoders in the presence of burst noise.

The complexity of GRAND-MO, defined as the maximum number of codebook membership queries done, is directly proportional to the number of putative test error patterns. In this paper, we propose a novel method for generating test error patterns to reduce the complexity of GRAND-MO decoding. Furthermore, we propose the first hardware architecture for GRAND-MO. Considering a code of length $128$ and a target FER of $10^{-5}$, the proposed architecture achieves an average throughput of $52$ Gbps, and outperforms GRANDAB \cite{GRANDAB-VLSI} by $3$ dB. As compared to the $(79,64)$ BCH code decoder \cite{Choi19}, the proposed VLSI architecture provides 33$\%$ higher worst-case throughput and a $2$ dB gain for a target FER of $10^{-5}$.

The rest of this paper is organized as follows: Section 2 describes the GRAND-MO algorithm and the channel model under consideration. Section 3 introduces complexity reduction techniques for GRAND-MO and their use to develop the proposed hardware architecture. Additionally, Section 3 presents a comparison of the proposed GRAND-MO architecture with GRANDAB and a newly developed BCH decoder. Finally, in Section 4, concluding remarks are made.

\section{Preliminaries}
\subsection{Notations}
Matrices are denoted by a bold upper-case letter ($\bm{M}$), while vectors are denoted with bold lower-case letters ($\bm{v}$).
The transpose operator is represented by $^\top$. 
The number of $k$-combinations from a given set of $n$ elements is noted by $\binom{n}{k}$.
$\mathds{1}_n$ is the indicator vector where all locations except the $n^{\text{th}}$ location are $0$ and the $n^{\text{th}}$ location is $1$. 
All the indices start at $1$.

\subsection{Channel Model}

In this work, the classic two-state Markov chain \cite{GilbertModel} is used to model a binary channel with burst noise. When the channel is in a good state, $G$, the channel is noiseless; however, when the channel is in a bad state, $B$, the channel becomes noisy and introduces errors. The transition probability from $G$ to $B$ is $b$, and the transition probability from $B$ to $G$ is $g$. Both $b$ and $g$ are assumed to be known and, in practice, can be estimated. A burst error is a sequence of consecutive errors introduced by the channel, with a length that follows a geometric distribution of mean $\frac{1}{g}$ and variance $\frac{1-g}{g^2}$. The Markov channel's stationary bit-flip probability $p$ is $\frac{b}{b+g}=\mathrm{Q}(\sqrt{2R\frac{E_{b}}{N_{0}}})$ where $R$ is the code rate. It should be noted that when $p=b$, the Markov channel transforms into a memoryless BSC.

\subsection{GRAND Markov Order}
Algorithm \ref{alg:grandMO}  summarizes GRAND-MO's  pseudo-code for a linear $(n,k)$ block code, where $n$ is the code length and $k$ is the number of information bits.
The algorithm's inputs are $\bm{r}$, $b$, $g$ and $\lfloor\frac{d}{2}\rfloor$ where $\bm{r}$ is the received vector  of size $n$ and $d$ is the minimum distance of the code. Moreover, the algorithm also utilizes the $(n-k)\times n$ parity check matrix $\bm{H}$ of the code  and the $n\times k$  matrix $\bm{G}^{-1}$, with $\bm{G}$ being the generator matrix of the code $(\bm{G}^{-1}\cdot\bm{G} = \bm{I})$. 

The error vector is initialized to $\bm{0}$ (line 1) in GRAND-MO, and ${\Delta}l$ is computed using $b$ and $g$ (line 2). Then, the test error patterns are generated sequentially by referring to ${\Delta}l$ (line 4). The generated error patterns have $m$ bursts and have a Hamming weight of $l$. Finally, $\bm{r}$ is combined with the current test error pattern, and the resulting word is queried for codebook membership by verifying that
\begin{equation}
\bm{H} \cdot(\bm{r} \oplus \bm{e})^\top
\label{eq:constraint}
\end{equation}
is equal to zero. If the resulting codeword belongs to the codebook, the message ($\hat{\bm{u}}$) is recovered (line 7). GRAND-MO decoding is terminated when the number of bursts $m$ in the generated test error pattern and the Hamming weight $l$ of the error pattern equal $\lfloor\frac{d}{2}\rfloor$.

\begin{algorithm}[t]
\caption{\label{alg:grandMO}GRAND Markov Order}
    \DontPrintSemicolon
    \SetAlgoVlined  
    \SetKwData{e}{$\bm{e}$}
    \SetKwData{d}{$\lfloor\frac{d}{2}\rfloor$}
    \SetKwData{b}{${b}$}
    \SetKwData{g}{${g}$}
    \SetKwData{ABANDON}{${ABANDON}$}
    \SetKwData{estm}{$\hat{\bm{u}}$}
    \SetKwData{ginv}{$\bm{G}^{-1}$}
    \KwIn{$\bm{H}$, \ginv, $\bm{r}$, \b, \g, \d}
    \KwOut{\estm $\text{OR}$ \ABANDON}
    \SetKwFunction{RecursiveComputeLLRs}{recursiveComputeLLRs}
    \SetKwFunction{DecodeRONE}{decodeR1}
    \SetKwFunction{RDecodeRONE}{redecodeR1}
    \SetKwFunction{DecodeRZERO}{decodeR0}
    \SetKwFunction{Find}{findCandidate}
    \SetKwFunction{new}{generateNewMarkovErrorPattern}
    $\e \leftarrow \bm{0}$\;
    ${\Delta}l \leftarrow \lfloor\frac{\log(\frac{b}{g})}{\log(\frac{1-g}{1-b})}\rfloor$\;
    \While{$\bm{H} \cdot(\bm{r} \oplus \e)^\top \neq \bm{0}$}{
        [\e, ${m}$, ${l}$] $\leftarrow$ \new{${\Delta}l$}\;
        \If{$m == \d$ $\text{AND}$ $l == \d$} {
            \KwRet{\ABANDON}
           }
    }
    $\estm \leftarrow (\bm{r} \oplus \e)\cdot\ginv$\;
    \KwRet{\estm}
\end{algorithm}

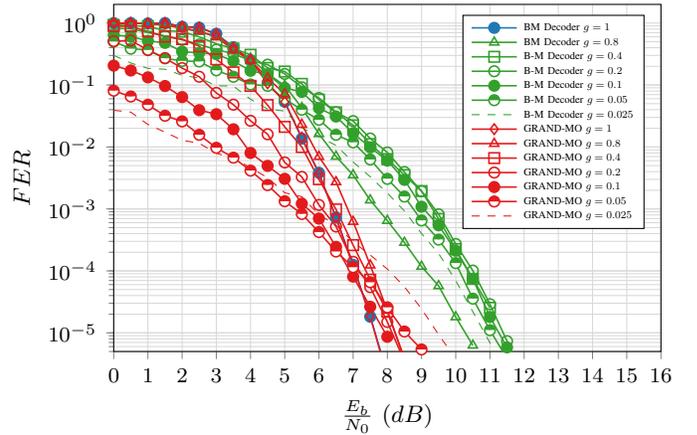
\begin{figure}[!t]
\begin{tikzpicture}
    \begin{semilogyaxis}[
            footnotesize, width=\columnwidth, height=.7\columnwidth,    
            xmin=0, xmax=16, xtick={0,1,...,16},
            ymin=0.5e-5,  ymax=2,
            xlabel=$\frac{E_{b}}{N_{0}}$ $(dB)$, ylabel=$FER$,  
            grid=both, grid style={gray!30},
            tick align=outside, tickpos=left, 
            legend pos=north east, legend cell align={left},
            legend style={nodes={scale=0.5, transform shape}},
            /pgfplots/table/ignore chars={|},
        ]

        \addplot[mark=* , Paired-1, semithick]  table[x=SNR, y=FER] {data/results/BCH/FER_BCH_GRANDAB_G1.txt};
        \addplot[mark=triangle, Paired-3, semithick]  table[x=SNR, y=FER] {data/results/BCH/FER_BCH_GRANDAB_G08.txt};
        \addplot[mark=square  , Paired-3, semithick]  table[x=SNR, y=FER] {data/results/BCH/FER_BCH_GRANDAB_G04.txt};
        \addplot[mark=o       , Paired-3, semithick]  table[x=SNR, y=FER] {data/results/BCH/FER_BCH_GRANDAB_G02.txt};
        \addplot[mark=* , Paired-3,  semithick]  table[x=SNR, y=FER] {data/results/BCH/FER_BCH_GRANDAB_G01.txt};
        \addplot[mark=halfcircle*, Paired-3,  semithick]  table[x=SNR, y=FER] {data/results/BCH/FER_BCH_GRANDAB_G005.txt};
        \addplot[mark=none , Paired-3,  dashed]  table[x=SNR, y=FER] {data/results/BCH/FER_BCH_GRANDAB_G0025.txt};
        \addplot[mark=diamond, Paired-5, semithick]  table[x=SNR, y=FER] {data/results/BCH/FER_BCH_GRANDMO_G1.txt};
        \addplot[mark=triangle, Paired-5, semithick]  table[x=SNR, y=FER] {data/results/BCH/FER_BCH_GRANDMO_G08.txt};
        \addplot[mark=square  , Paired-5, semithick]  table[x=SNR, y=FER] {data/results/BCH/FER_BCH_GRANDMO_G04.txt};
        \addplot[mark=o       , Paired-5, semithick]  table[x=SNR, y=FER] {data/results/BCH/FER_BCH_GRANDMO_G02.txt};
        \addplot[mark=* , Paired-5,  semithick]  table[x=SNR, y=FER] {data/results/BCH/FER_BCH_GRANDMO_G01.txt};
        \addplot[mark=halfcircle*, Paired-5,  semithick]  table[x=SNR, y=FER] {data/results/BCH/FER_BCH_GRANDMO_G005.txt};
        \addplot[mark=none  , Paired-5,  dashed]  table[x=SNR, y=FER] {data/results/BCH/FER_BCH_GRANDMO_G0025.txt};
         \legend{{} {BM Decoder $g=1$},{} {BM Decoder $g=0.8$},{} {B-M Decoder $g=0.4$},{} {B-M Decoder $g=0.2$},
         {} {B-M Decoder $g=0.1$},{} {B-M Decoder $g=0.05$},{} {B-M Decoder $g=0.025$},
         {} GRAND-MO $g=1$,{} {GRAND-MO $g=0.8$},{} {GRAND-MO $g=0.4$},{} {GRAND-MO $g=0.2$},
         {} {GRAND-MO $g=0.1$},{} {GRAND-MO $g=0.05$},{} {GRAND-MO $g=0.025$},
         }
    \end{semilogyaxis}
\end{tikzpicture}  
\caption{\label{fig:fer_bch}Comparison of the GRAND-MO and BCH Berlekamp-Massey (B-M) decoding performance of BCH code (127, 106) in Markov channels.}
\end{figure}
\begin{figure}[!t]
\begin{tikzpicture}
    \begin{semilogyaxis}[
            footnotesize, width=\columnwidth, height=.7\columnwidth,    
            xmin=0, xmax=18, xtick={0,1,...,18},
            ymin=1e-5,  ymax=2,
            xlabel=$\frac{E_{b}}{N_{0}}$ $(dB)$, ylabel=$FER$,  
            grid=both, grid style={gray!30},
            tick align=outside, tickpos=left, 
            legend pos=north east, legend cell align={left},
            legend style={nodes={scale=0.6, transform shape}},
            /pgfplots/table/ignore chars={|},
        ]

        \addplot[mark=triangle, Paired-3, semithick]  table[x=SNR, y=FER] {data/results/RLC/FER_RLC_GRANDAB_G08.txt};
        \addplot[mark=square  , Paired-3, semithick]  table[x=SNR, y=FER] {data/results/RLC/FER_RLC_GRANDAB_G04.txt};
        \addplot[mark=o       , Paired-3, semithick]  table[x=SNR, y=FER] {data/results/RLC/FER_RLC_GRANDAB_G02.txt};
        \addplot[mark=triangle, Paired-5, semithick]  table[x=SNR, y=FER] {data/results/RLC/FER_RLC_GRANDMO_G08.txt};
        \addplot[mark=square  , Paired-5, semithick]  table[x=SNR, y=FER] {data/results/RLC/FER_RLC_GRANDMO_G04.txt};
        \addplot[mark=o       , Paired-5, semithick]  table[x=SNR, y=FER] {data/results/RLC/FER_RLC_GRANDMO_G02.txt};
        \addplot[mark=triangle, Paired-9, semithick]  table[x=SNR, y=FER] {data/results/RLC/FER_RLC_GRANDMO_G08_32_08.txt};
        \addplot[mark=square  , Paired-9, semithick]  table[x=SNR, y=FER] {data/results/RLC/FER_RLC_GRANDMO_G04_32_16.txt};
        \addplot[mark=o       , Paired-9, semithick]  table[x=SNR, y=FER] {data/results/RLC/FER_RLC_GRANDMO_G02_32_24.txt};
        \legend{{} {GRANDAB ($AB=3$, $g=0.8$)  },
            {} {GRANDAB ($AB=3$, $g=0.4$)  },
            {} {GRANDAB ($AB=3$, $g=0.2$)  },
         {} {GRAND-MO ($g=0.8$)},{} {GRAND-MO  ($g=0.4$)},{} {GRAND-MO  ($g=0.2$)},
         {} {$g=0.8$, $m=2$, $l_1=32$, $l_2=8$},{} {$g=0.4$, $m=2$, $l_1=32$, $l_2=16$},{} {$g=0.2$, $m=2$, $l_1=32$, $l_2=24$},
         }
    \end{semilogyaxis}
\end{tikzpicture}  
\caption{\label{fig:fer_RLC}Comparison of the GRANDAB ($AB=3$) and GRAND-MO decoding performance using RLC code (128, 104) with Markov query order and proposed query order ($g$, $m$, $l_1$, $l_2$).}
\end{figure}
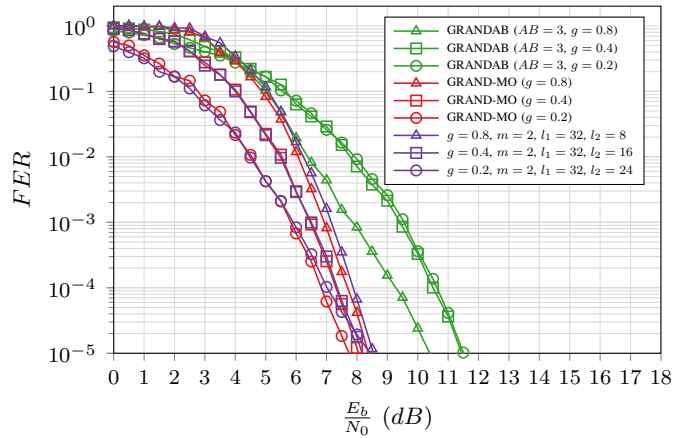
\begin{figure}
  \centering
  \includegraphics[width=1\linewidth]{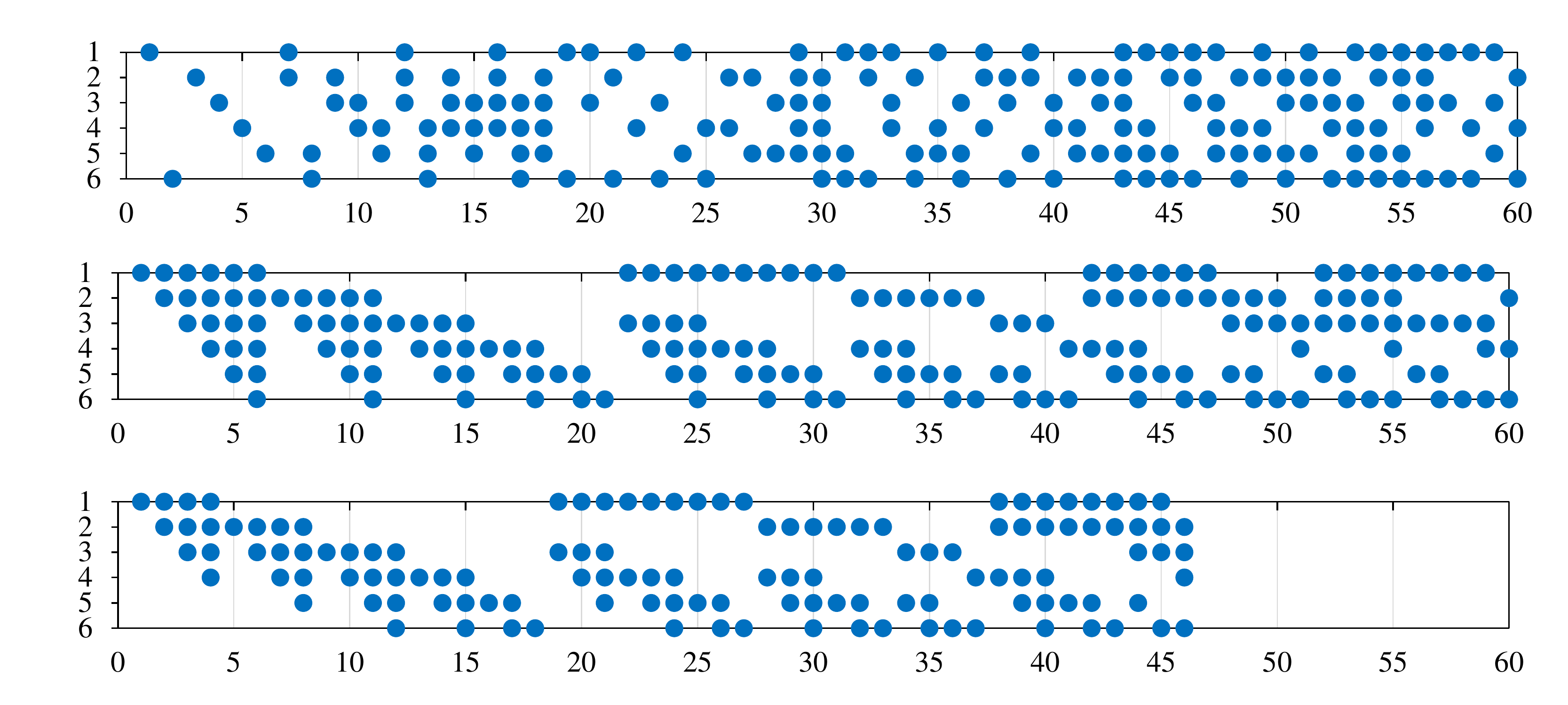}
  \caption{Test error pattern generation for GRAND-MO for  $n=6$ and ${\Delta}l=2$  (a) Upper: Markov query order (b) Middle: Proposed re-arranged query order (c) Bottom: Proposed query order with parameters ($m=2$, $l_1=4$ and $l_2=3$).}
  \label{fig:garndMO_SCH} 
\end{figure}
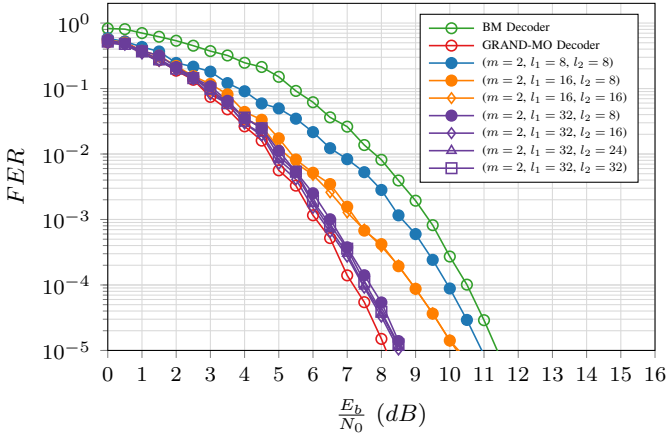
\begin{figure}[!t]
\begin{tikzpicture}
    \begin{semilogyaxis}[
            footnotesize, width=\columnwidth, height=.7\columnwidth,    
            xmin=0, xmax=16, xtick={0,1,...,16},
            ymin=1e-5,  ymax=2,
            xlabel=$\frac{E_{b}}{N_{0}}$ $(dB)$, ylabel=$FER$,  
            grid=both, grid style={gray!30},
            tick align=outside, tickpos=left, 
            legend pos=north east, legend cell align={left},
            legend style={nodes={scale=0.6, transform shape}},
            /pgfplots/table/ignore chars={|},
        ]

         \addplot[mark=o       , Paired-3, semithick]  table[x=SNR, y=FER] {data/results/BCH/FER_BCH_GRANDAB_G02.txt};
         \addplot[mark=o       , Paired-5, semithick]  table[x=SNR, y=FER] {data/results/BCH/FER_BCH_GRANDMO_G02.txt};
         \addplot[mark=otimes*,       Paired-1, semithick]  table[x=SNR, y=FER] {data/results/BCH/FER_BCH_GRANDMO_G02_8_8.txt};
         \addplot[mark=otimes*,  Paired-7, semithick]  table[x=SNR, y=FER] {data/results/BCH/FER_BCH_GRANDMO_G02_16_8.txt};
         \addplot[mark=diamond, Paired-7, semithick]  table[x=SNR, y=FER] {data/results/BCH/FER_BCH_GRANDMO_G02_16_16.txt};
         \addplot[mark=otimes*  , Paired-9, semithick]  table[x=SNR, y=FER] {data/results/BCH/FER_BCH_GRANDMO_G02_32_8.txt};
         \addplot[mark=diamond  , Paired-9, semithick]  table[x=SNR, y=FER] {data/results/BCH/FER_BCH_GRANDMO_G02_32_16.txt};
         \addplot[mark=triangle  , Paired-9, semithick]  table[x=SNR, y=FER] {data/results/BCH/FER_BCH_GRANDMO_G02_32_24.txt};
         \addplot[mark=square  , Paired-9, semithick,]  table[x=SNR, y=FER] {data/results/BCH/FER_BCH_GRANDMO_G02_32_32.txt};
         \legend{{} {BM Decoder },
         {} {GRAND-MO Decoder},
         {} {($m=2$, $l_1=8$, $l_2=8$)},{} {($m=2$, $l_1=16$, $l_2=8$)},
         {} {($m=2$, $l_1=16$, $l_2=16$)},{} {($m=2$, $l_1=32$, $l_2=8$)},
         {} {($m=2$, $l_1=32$, $l_2=16$)},{} {($m=2$, $l_1=32$, $l_2=24$)},
         {} {($m=2$, $l_1=32$, $l_2=32$)},
         }
    \end{semilogyaxis}
\end{tikzpicture}  
\caption{\label{fig:fer_bch_pro}Comparison of the GRAND-MO ($g=0.2$) decoding performance using BCH code (127, 106) with Markov query order and proposed query order with parameters ($m$, $l_1$, $l_2$).}
\end{figure}
The frame error rate (FER) performance for GRAND-MO decoding of BCH code (127, 106) in Markov channels is plotted in Fig. \ref{fig:fer_bch}. The demodulator provides hard decision values to the GRAND-MO decoder. It is noted that GRAND-MO's performance improves as channel memory increases ($g$ decreases), while the traditional BCH Berlekamp-Massey (B-M) decoder \cite{Berlekamp68,Massey69} shows a degradation in FER performance with the increase in channel memory. GRAND-MO's performance differs from that of the BCH decoder because GRAND-MO adjusts its error pattern generation to mitigate the impact of noise bursts in the channel.

Similar trends can be observed with Random Linear Codes (RLCs). RLCs are linear block codes that are theoretically known to be high-performing \cite{RLC1,RLC2}, but not considered practical in terms of decodability. Fig. \ref{fig:fer_RLC} plots the FER performance for GRAND-MO and GRANDAB \cite{Duffy19TIT} decoding of RLCs of length $n=128$. We can observe that with the decrease in $g$, GRAND-MO outperforms GRANDAB (AB = 3) decoder in FER performance.

\section{VLSI Architecture for GRAND-MO}
We describe the proposed VLSI architecture for GRAND-MO decoding in this section. Furthermore, we analyse the error patterns generated by GRAND-MO and suggest simplifications to the test pattern generation process.

\subsection{Test error pattern generation for GRAND-MO}
GRAND-MO generates test error patterns in Markov query order. Fig. \ref{fig:garndMO_SCH} (a) depicts the Markov query order for code length $n=6$ and ${\Delta}l=2$ where each column corresponds to a putative error pattern and a dot corresponds to a flipped bit location. As presented in Fig. \ref{fig:garndMO_SCH} (b) and discussed in section \ref{s:Principle}, we propose rearranging these error patterns to simplify the hardware implementation.

The maximum number of codebook membership queries (and hence the worst-case complexity) for GRAND-MO decoding is determined by ${\Delta}l$, $\frac{1}{g}$, and $\frac{E_{b}}{N_{0}}$. It should be noted that the average number of codebook membership queries for GRAND and its variants is far lower than the maximum number of codebook membership queries. To reduce worst-case complexity, we suggest restricting the number of bursts $m$ as well as burst sizes $l_m$ for the generated test error patterns. Figure \ref{fig:garndMO_SCH} (c) depicts the generation of a modified test error pattern with parameters $m=2$, $l_1=4$ and $l_2=3$. In comparison to the Markov query order, the proposed query order with parameters ($m=2$,\text{ }  $l_1=4$ and $l_2=3$) reduces the worst-case complexity from $\numprint{60}$ to $\numprint{46}$ test error patterns. 

The FER performance for GRAND-MO decoding of BCH code (127, 106) with $g=0.2$ is shown in Fig. \ref{fig:fer_bch_pro}. The performance of the decoder using the proposed query order with different parameters $(m,\text{ } l_m)$, is compared to the performance of the decoder with the original Markov query order. The proposed query order with parameters $(m=2,\text{ }  l_1=32,\text{ }  l_2=8)$ results in a $0.3$ \text{dB} degradation in FER at $10^{-5}$; however, the maximum number of codebook membership queries is reduced from  $\numprint{3530504}$ queries required by Markov order to $\numprint{487818}$ queries at $\frac{E_b}{N_0}=8$ \text{dB}. Similarly, for RLCs, the proposed query order's parameters $m$ and $l_m$ can be adjusted to match the FER performance of GRAND-MO with Markov query order for different lengths, rates, and average burst lengths. Fig. \ref{fig:fer_RLC} shows the FER performance for GRAND-MO decoding with the Markov query order and the proposed query order for RLC code (128, 104).

\begin{figure}
  \centering
  \includegraphics[width=1\linewidth]{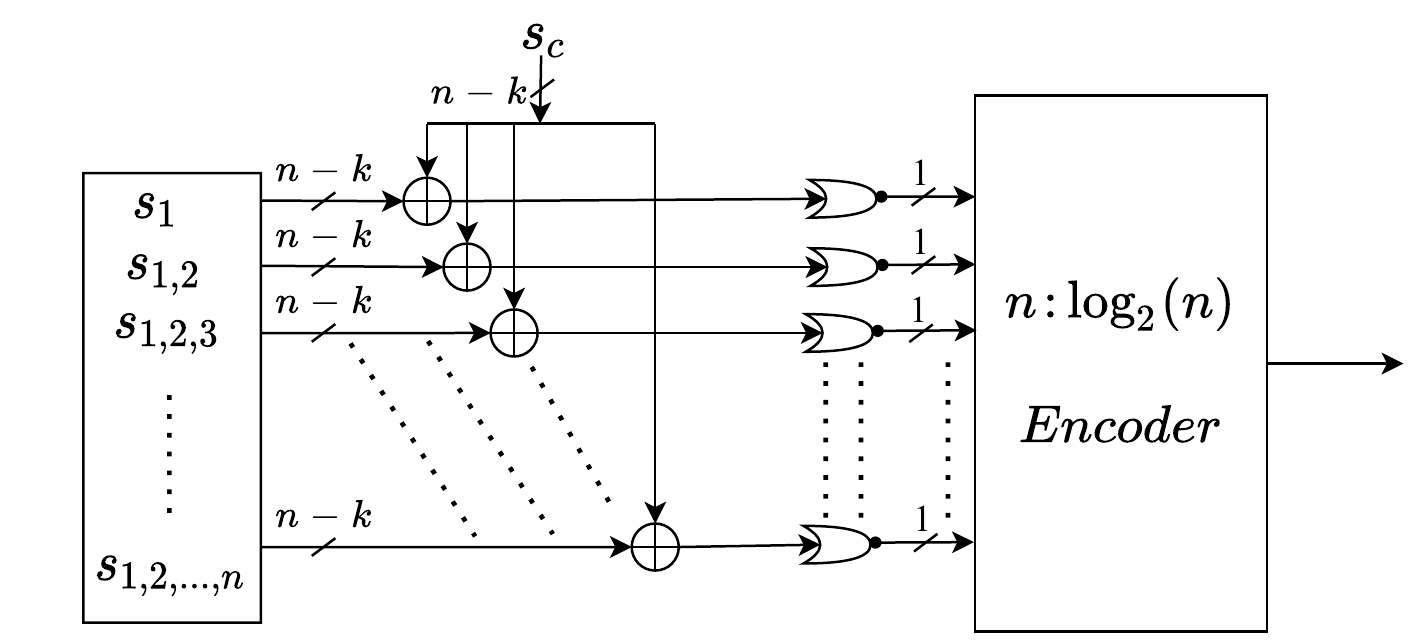}
  \caption{Checking test error patterns corresponding to a noise burst of length $1\leq{l}\leq{n}$, where $\bm{s}_{1,2\ldots,l} = \bm{H}\cdot\mathds{1}_1^\top \oplus \bm{H}\cdot\mathds{1}_2^\top \ldots \oplus\bm{H}\cdot\mathds{1}_l^\top$. } 
  \label{fig:VLSI_1} 
\end{figure}
\begin{figure}[!t]
    \captionsetup[subfloat]{farskip=0pt}
    \centering
    \subfloat[Checking all test error patterns corresponding to a single noise burst of size $l\leq4$ in one time-step]
    {
        \includegraphics[width=1\linewidth]{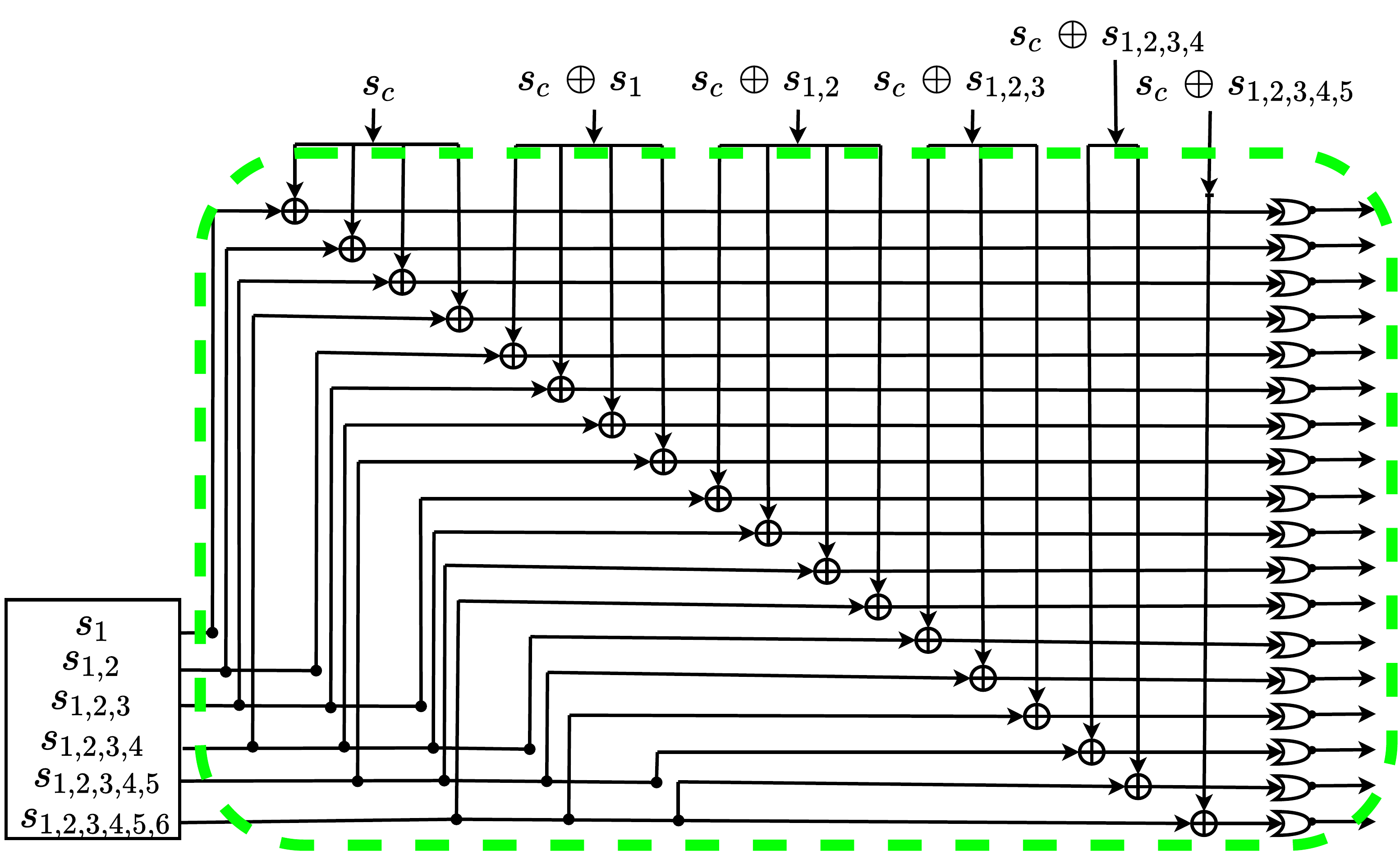}
        \label{fig:VLSI_2:a}
    }\hfil
    \subfloat[Error patterns corresponding to a single noise burst of size $l\leq4$ (red-rectangle).]
    {
        \includegraphics[width=1\linewidth]{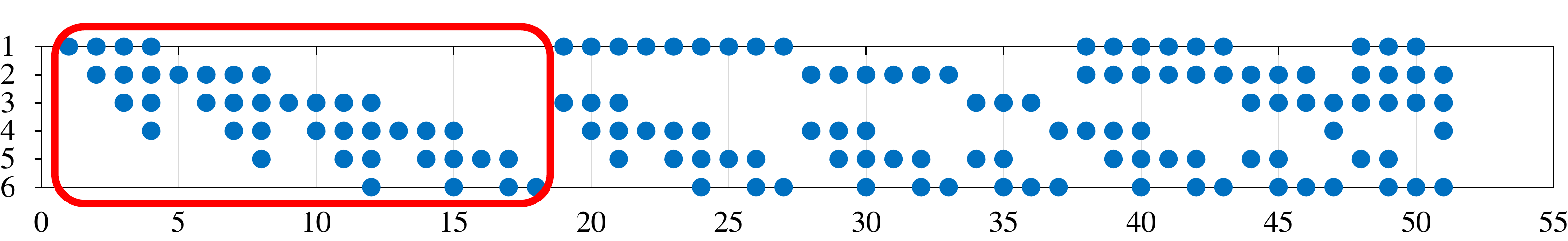}
        \label{fig:VLSI_2:b}
    }
    \caption{\label{fig:VLSI_2} Checking test error patterns for GRAND-MO decoding corresponding to proposed query order with parameters ($n=6$, $m=2$, $l_1=4$ and $l_2=3$).}
\end{figure}
\begin{figure}[!t]
    \captionsetup[subfloat]{farskip=0pt}
    \centering
    \subfloat[Checking all test error patterns corresponding to $m=2$ and $l_1=1$ with $\bm{s}_{comp}=s_c\oplus{s_1}$.]
    {
        \includegraphics[width=1\linewidth]{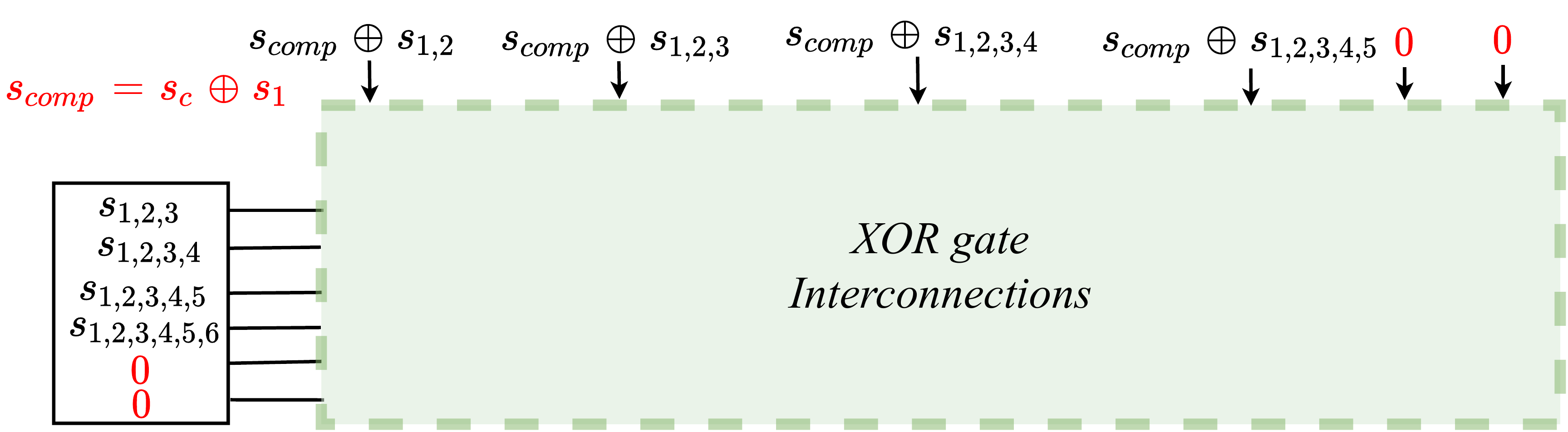}
        \label{fig:VLSI_3:a}
    }\hfil
    \subfloat[Error patterns corresponding to $m=2$ and $l_1=1$ with $\bm{s}_{comp}=s_c\oplus{s_1}$ and $n=6$ (red-rectangle).]
    {
        \includegraphics[width=1\linewidth]{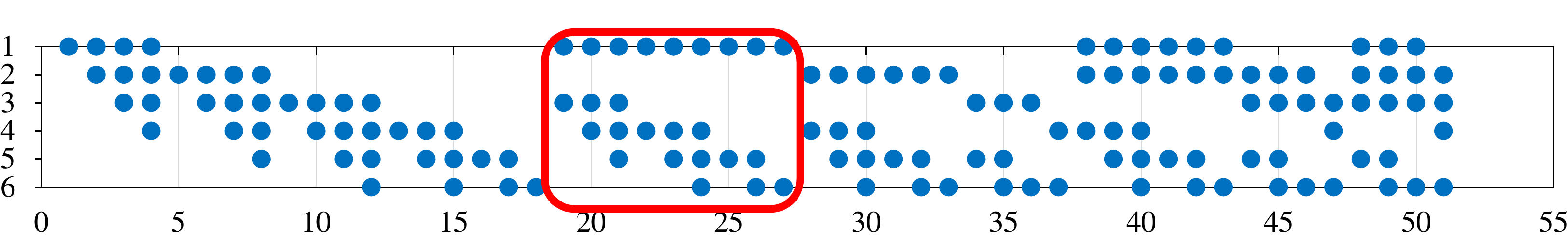}
        \label{fig:VLSI_3:b}
    }
    \caption{\label{fig:VLSI_3} Checking test error patterns corresponding to proposed query order for GRAND-MO at time step 2.}
\end{figure}
\subsection{Principle, Scheduling and Details}
\label{s:Principle}
For a $(n,k)$ linear block code, a VLSI architecture for GRANDAB (AB=3) decoder was proposed in \cite{GRANDAB-VLSI}. The proposed architecture uses $n\times (n-k)\text{-bit}$ shift registers to store $(n-k)\text{-bit}$ syndromes of 1-bit flip error patterns ($\bm{s}_i = \bm{H}\cdot\mathds{1}_i^\top$ with $i \in \llbracket 1\isep n \rrbracket$). Moreover, the proposed decoder uses the linearity property of the underlying code to combine $l$  1-bit flip error syndrome to generate an error pattern with the Hamming weight of $l$ ($\bm{s}_{1,2\ldots,l} = \bm{H}\cdot\mathds{1}_1^\top \oplus \bm{H}\cdot\mathds{1}_2^\top \ldots \oplus\bm{H}\cdot\mathds{1}_l^\top $). By shifting the data stored in the shift registers, error pattern syndromes corresponding to different bit flip patterns are generated. This approach forms the basis for the proposed GRAND-MO architecture. Since the proposed architecture for GRANDAB \cite{GRANDAB-VLSI}  can only generate test error patterns with Hamming weights $\leq3$, significant improvements are needed to support generating error patterns with burst lengths $l\geq3$.
\begin{figure}[!t]
    \captionsetup[subfloat]{farskip=0pt}
    \centering
    \subfloat[Checking all test error patterns corresponding corresponding to $m=2$ and $l_1=1$ with $\bm{s}_{comp}=s_c\oplus{s_2}$.]
    {
        \includegraphics[width=1\linewidth]{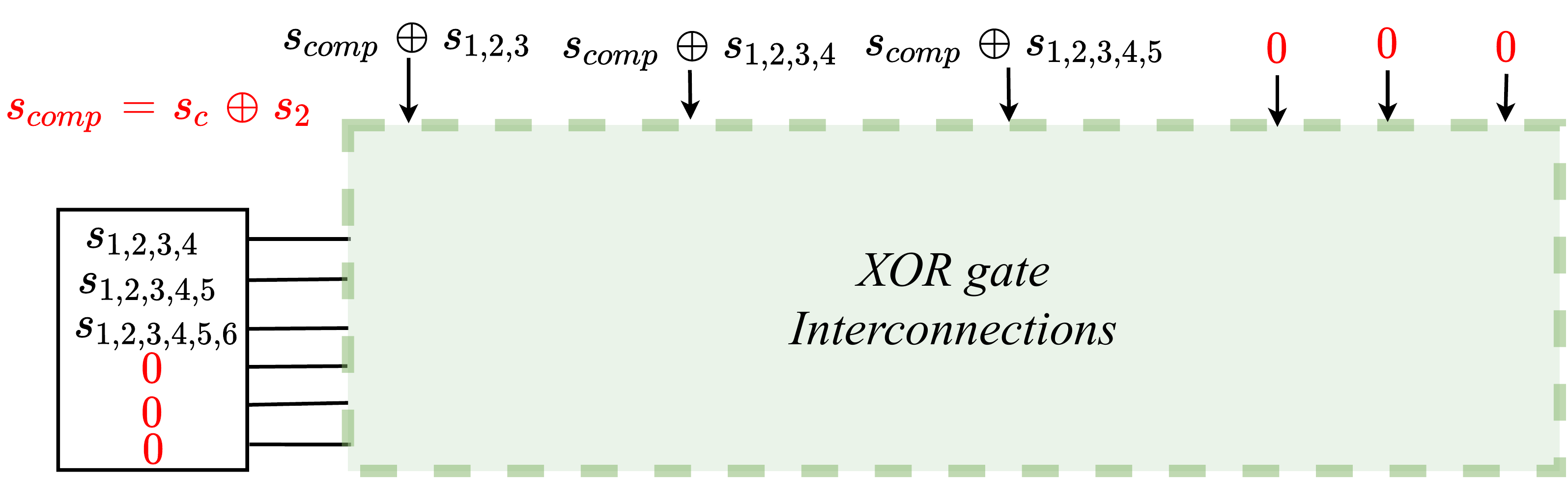}
        \label{fig:VLSI_4:a}
    }\hfil
    \subfloat[Error patterns corresponding to $m=2$ and $l_1=1$ with $\bm{s}_{comp}=s_c\oplus{s_2}$ and $n=6$ (red-rectangle).]
    {
        \includegraphics[width=1\linewidth]{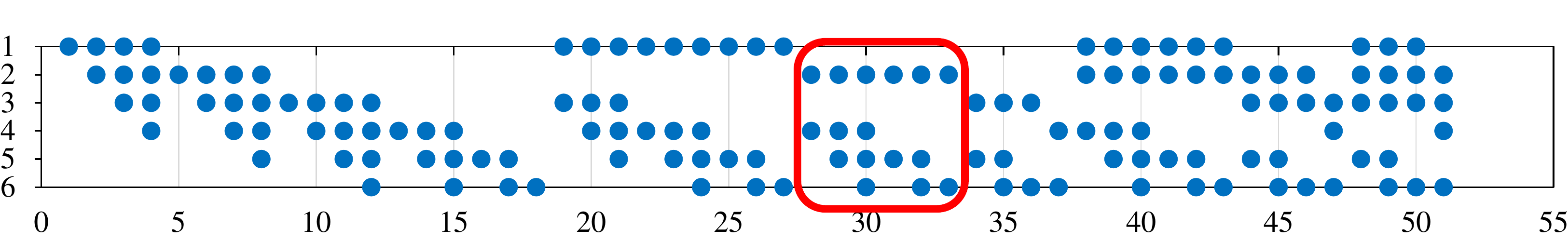}
        \label{fig:VLSI_4:b}
    }
    \caption{\label{fig:VLSI_4} Checking test error patterns corresponding to proposed query order for GRAND-MO at time step 3.}
\end{figure}
\begin{figure}[!t]
    \captionsetup[subfloat]{farskip=0pt}
    \centering
    \subfloat[Checking all test error patterns corresponding to $m=2$ and $l_1=2$ with $\bm{s}_{comp}=s_c\oplus{s_{1,2}}$.]
    {
        \includegraphics[width=1\linewidth]{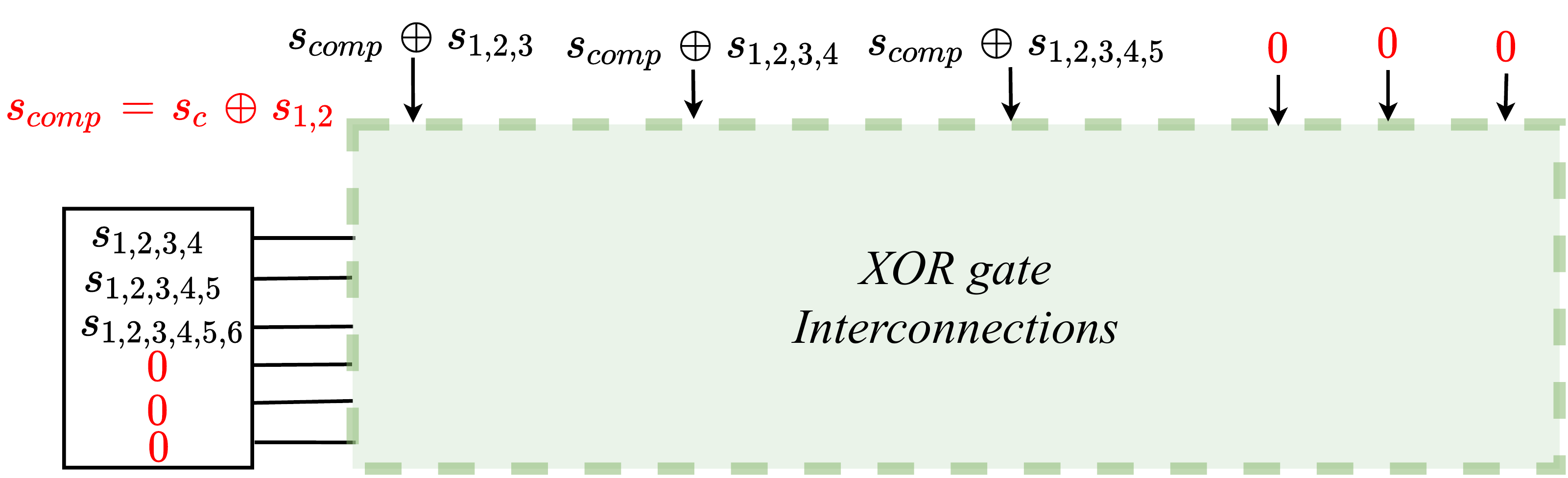}
        \label{fig:VLSI_5:a}
    }\hfil
    \subfloat[Error patterns corresponding to $m=2$ and $l_1=2$ with $\bm{s}_{comp}=s_c\oplus{s_{1,2}}$ and $n=6$ (red-rectangle).]
    {
        \includegraphics[width=1\linewidth]{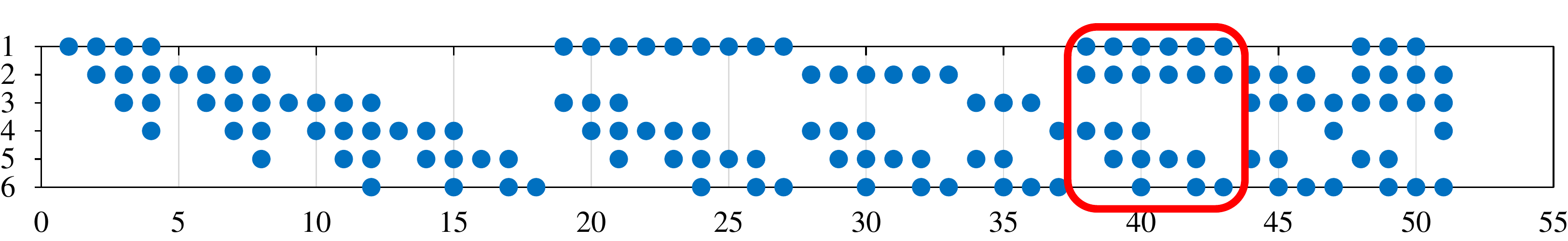}
        \label{fig:VLSI_5:b}
    }
    \caption{\label{fig:VLSI_5} Checking test error patterns corresponding to proposed query order for GRAND-MO at time step 6.}
\end{figure}
\begin{figure*}
\centering
  \includegraphics[width=0.8\linewidth]{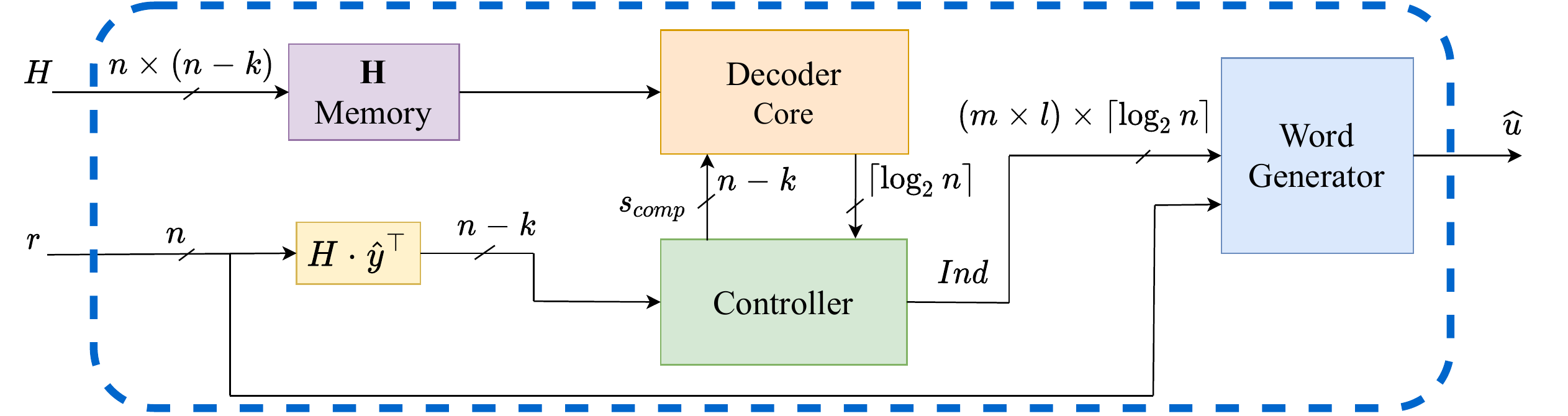}
  \caption{Proposed VLSI Architecture for GRAND-MO.}
  \label{fig:arch}
\vspace{-1em}
\end{figure*}

Fig. \ref{fig:VLSI_1} presents the contents of the $n\times (n-k)\text{-bit}$ shift register and the associated peripheral circuitry. This structure is used to generate the test error patterns corresponding to a noise burst of length $l$ $(1\leq{l}\leq{n})$. Each row of the shift register stores a syndrome corresponding to a noise burst, such that the $l^{th}$ row stores the syndrome corresponding to a noise burst of length $l$ ($\bm{s}_{1,2\ldots,l} = \bm{H}\cdot\mathds{1}_1^\top \oplus \bm{H}\cdot\mathds{1}_2^\top \ldots \oplus\bm{H}\cdot\mathds{1}_l^\top $). Through combining each row of the shift register with the syndrome of the received vector $s_c$ ($s_c=\bm{H}\cdot\bm{r}^\top$) using the $(n-k)$-bit-wide XOR gates, we can compute the syndrome of the test error patterns corresponding to a noise burst of length $l$. Each of the $n$ test syndromes is NOR-reduced, to feed an $n$-to-$\log_2n$ priority encoder. The output of each NOR-reduce is 1 if and only if all the bits of the syndrome computed by (\ref{eq:constraint}) are 0.

Based on the example presented in Fig. \ref{fig:garndMO_SCH} (c), we explain the VLSI architecture for the proposed query order for GRAND-MO decoding. An example of the contents of the shift register and the arrangement of XOR gates is presented in Fig. \ref{fig:VLSI_2}. This structure is used to generate test error patterns corresponding to the proposed query order with parameters $n=6$, $m=2$, $l_1=4$ and $l_2=3$. For the sake of clarity, the priority encoder and its associated signals are omitted in the figure. Due to the use of a specific arrangement of shift register and XOR gates, all the error patterns corresponding to a single noise burst ($m=1$) of size $l\leq4$ are checked (\ref{eq:constraint}) in a single time step as presented in Fig. \ref{fig:VLSI_2} (b).

To generate the test error patterns corresponding to $m>1$, a controller is used in conjunction with the shift register. Fig. \ref{fig:VLSI_3} shows the contents of the shift register and the syndrome that is outputted by the controller, which is denoted as $\bm{s}_{comp}$, to 
generate test error patterns corresponding to $m=2$ and $l_1=1$. The shift register is shifted-up by 2 positions and the controller outputs $\bm{s}_{comp}=s_c\oplus{s_1}$. Hence, all the test error patterns with $\bm{s}_{comp}=s_c\oplus{s_1}$ are checked in one time step. At the next time step, the controller outputs $\bm{s}_{comp}=s_c\oplus{s_2}$ and the shift register is shifted up by 1 position. This allows us to generate all  the test error patterns, with $\bm{s}_{comp}=s_c\oplus{s_2}$ as shown in Fig. \ref{fig:VLSI_4}. Therefore, for a code length of $n$, $n-2$ time steps are required to generate all test error patterns corresponding to $m=2$ and $l_1=1$ where shift register is shifted up by 1 in each time step.

Similarly, to generate test error patterns corresponding to $m=2$ and $l_1=2$, the shift register is reset and shifted-up by 3 positions. In this position, the controller outputs $\bm{s}_{comp}=s_c\oplus{s_1}\oplus{s_2}$ as shown in Fig. \ref{fig:VLSI_5}. A total number of $n-3$ time steps are required to generate all test error patterns corresponding to $m=2$ and $l_1=2$ since the shift register is shifted up by 1 in each time step. In summary, for the proposed VLSI architecture, the number of required time steps to check all the error patterns corresponding to the proposed query order with parameters $(n,m=2,L=min(l_1,l_2)])$ is
given by: 
\begin{equation}
\label{eq:nb_steps}
L\times(\frac{2\times{n}-L-3}{2})+2.
\end{equation}

The proposed VLSI architecture for GRAND-MO is presented in Fig.~\ref{fig:arch}. For clarity, the control and clock signals are not shown. The proposed architecture takes $\bm{r}$ as input and generates the estimated word as output $\hat{\bm{u}}$. At any time, to support any code, given the length and rate constraints, an $\bm{H}$ matrix can be loaded into the \textit{H memory}. To begin, a syndrome check is performed on $\bm{r}$ to determine whether the received vector is a valid codeword. If the syndrome is verified, decoding is assumed to be successful and we terminate by outputting $\hat{\bm{u}}=\bm{r}$. Otherwise, the \textit{decoding core} applies test error patterns in the proposed query order until one of the test error pattern verifies the parity check constraint (\ref{eq:constraint}). After verifying that the resulting codeword belongs to the codebook, the  \textit{controller} module forwards the respective indices to the word generator module which translates these index values to their correct bit flip locations.
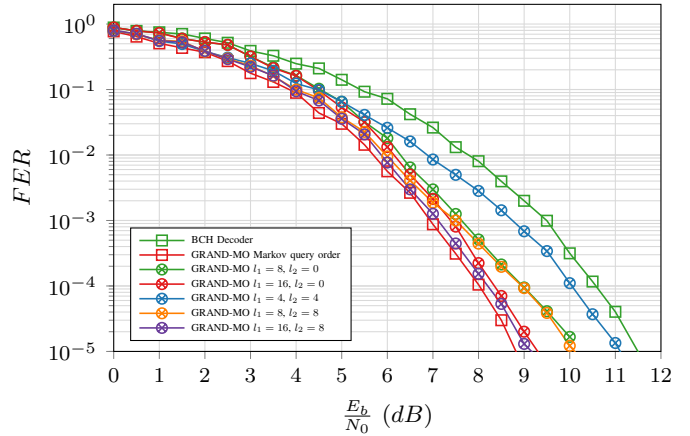
\begin{figure}[!t]
\begin{tikzpicture}
    \begin{semilogyaxis}[
            footnotesize, width=\columnwidth, height=.7\columnwidth,    
            xmin=0, xmax=12, xtick={0,1,...,12},
            ymin=1e-5,  ymax=2,
            xlabel=$\frac{E_{b}}{N_{0}}$ $(dB)$, ylabel=$FER$,  
            grid=both, grid style={gray!30},
            tick align=outside, tickpos=left, 
           legend pos=south west, legend cell align={left},
            legend style={nodes={scale=0.5, transform shape}},
            /pgfplots/table/ignore chars={|},
        ]

        \addplot[mark=square  , Paired-3, semithick]  table[x=SNR, y=FER] {data/results/BCH_79_64/FER_BCH_GRANDAB_G_04.txt};

        \addplot[mark=square  , Paired-5, semithick]  table[x=SNR, y=FER] {data/results/BCH_79_64/FER_BCH_GRANDMO_G_04.txt};
       \addplot[mark=otimes, Paired-3, semithick]  table[x=SNR, y=FER] {data/results/BCH_79_64/FER_BCH_GRANDMO_G_04_8_0.txt};
       \addplot[mark=otimes, Paired-5, semithick]  table[x=SNR, y=FER] {data/results/BCH_79_64/FER_BCH_GRANDMO_G_04_16_0.txt};
       \addplot[mark=otimes, Paired-1, semithick]  table[x=SNR, y=FER] {data/results/BCH_79_64/FER_BCH_GRANDMO_G_04_4_4.txt};
        \addplot[mark=otimes, Paired-7, semithick]  table[x=SNR, y=FER] {data/results/BCH_79_64/FER_BCH_GRANDMO_G_04_8_8.txt};
        \addplot[mark=otimes, Paired-9, semithick]  table[x=SNR, y=FER] {data/results/BCH_79_64/FER_BCH_GRANDMO_G_04_16_8.txt};
        \legend{{} {BCH Decoder }, 
          {} {GRAND-MO Markov query order }, {} {GRAND-MO $l_1=8$, $l_2=0$}, {} {GRAND-MO $l_1=16$, $l_2=0$},
         {} {GRAND-MO $l_1=4$, $l_2=4$},{} {GRAND-MO $l_1=8$, $l_2=8$}, {} {GRAND-MO $l_1=16$, $l_2=8$}, 
         }

    \end{semilogyaxis}
\end{tikzpicture}  
\caption{\label{fig:fer_bch_79} Comparison of the GRAND-MO decoding and BCH (PGZ) decoding performance for BCH code (79, 64) in Markov channels ($g=0.4$).}
\end{figure}
\subsection{Implementation Results}
The proposed GRAND-MO architecture with parameters ($m=2$, $l_1\leq32$ and $l_2\leq32$), has been implemented in Verilog HDL and synthesized using Synopsys Design Compiler with general-purpose TSMC 65 nm CMOS technology. The design has been verified using test benches generated via the bit-true C model of the proposed hardware. Table \ref{table:tableGRANDMO} presents the synthesis results for the proposed decoder with $n=128$, code rates between $0.75$ and $1$. 

The GRAND-MO implementation can support a maximum frequency of $500~\text{MHz}$. Since no pipelining strategy is used, one clock cycle corresponds to one time-step. For $n= 128$ and parameters ($m=2$, $l_1\leq32$ and $l_2\leq32$) $\numprint{3538}$ cycles (\ref{eq:nb_steps}) are required in the worst-case (W.C.) scenario, resulting in a W.C. latency of 7.0 $\mu$s. The average latency, however, is only 2 $ns$ at target FER of $10^{-5}$, which results in an average decoding information throughput of $52$ Gbps for the (128,104) RLC code presented in Fig. \ref{fig:fer_RLC}. The proposed GRAND-MO decoder has a $2.8\times$ area overhead as compared to the hard decision-based GRANDAB decoder (AB=3) \cite{GRANDAB-VLSI}. The average decoding throughput for both the proposed GRANDMO and GRANDAB decoder \cite{GRANDAB-VLSI} is comparable. The proposed GRAND-MO decoder, on the other hand, has $13.6\%$ lower W.C. latency, resulting in $13.6\%$  higher W.C. decoding throughput. Furthermore, as seen in Fig. \ref{fig:fer_RLC}. GRAND-MO's decoding performance with parameters ($g=0.4$, $m=2$, $l_1=32$, and $l_2=16$) outperforms GRANDAB decoder by at least $3$ dB for target FERs less than $10^{-5}$.

Recently, a high throughput VLSI architecture for a $(79,64)$ BCH code decoder based on the Peterson-Gorenstein–Zierler (PGZ) algorithm \cite{Peterson60} was proposed in \cite{Choi19}. The decoding performance of the (79,64) BCH PGZ decoder is compared with that of GRAND-MO decoding in Fig. \ref{fig:fer_bch_79}. As seen in Fig. \ref{fig:fer_bch_79}, GRAND-MO with proposed query order and parameters ($m=1$, $l_1=16$ and $l_2=0$) outperforms BCH decoder by $2$ dB for target FER of $10^{-5}$

 Table \ref{table:synth_79} compares the implementation results for GRAND-MO ($m=1$, $l_1\leq16$, and $l_2=0$) and the BCH decoder in \cite{Choi19}. For $n= 79$ and ($m=1$, $l_1=16$, and $l_2=0$), $\numprint{2}$ cycles (\ref{eq:nb_steps}) are required in the W.C. scenario. Even though the proposed decoder is $69\times$ bigger than the PGZ decoder in \cite{Choi19}, the W.C. latency is reduced by $33\%$ resulting in a W.C throughput of 32 Gbps. For target FER of $10^{-5}$, the proposed decoder exhibits a slightly better minimum latency and achieves an information throughput of 64 Gbps, while the BCH decoder is limited to 58 Gbps. In addition, the proposed GRAND-MO architecture can decode any code with $n=79$ and $R\geq0.75$, while \cite{Choi19} can only decode the (79,64) BCH code.

\begin{table}[t]
\centering
\caption{\label{table:tableGRANDMO}TSMC 65 nm CMOS Implementation Comparison for GRANDAB with GRAND-MO for $n=128$.}
\begin{adjustbox}{max width=\columnwidth}
\begin{tabular}{lrr}
\toprule
                             & GRANDAB \cite{GRANDAB-VLSI} & GRAND-MO                      \\
                            \cmidrule(l){2-2}\cmidrule(l){3-3}
Parameters                   & $AB=3$    & $m=2$, $l_1\leq32$ and $l_2\leq32$  \\  
Technology (nm)              & 65      & 65                    \\
Supply (V)                   & 0.9     & 0.9                   \\
Max. Frequency (MHz)         & 500     & 500                   \\
Area (mm\textsuperscript{2}) & 0.25    & 0.71                  \\
W.C. Latency (ns)            & 8196    & 7076                 \\
Avg. Latency (ns)            & 2       & 2                    \\
W.C. T/P (Mbps)              & 12.68    & 14.69                  \\
Avg. T/P (Gbps)              & 52    & 52                  \\
Code compatible              & Yes     & Yes                   \\
 Rate compatible              & Yes     & Yes                  \\
\bottomrule
\end{tabular}
\end{adjustbox}
\end{table}

\begin{table}[!t]
\centering
\caption{\label{table:synth_79}TSMC 65 nm CMOS Implementation Comparison for BCH decoder with GRAND-MO ($m=1$, $l_1\leq16$ and $l_2=0$) for $n=79$.}
\begin{adjustbox}{max width=\columnwidth}
\begin{tabular}{@{}lrrr@{}}
\toprule
                              & GRAND-MO      & (79,64) BCH decoder \cite{Choi19}     \\
                              \cmidrule(l){2-2}\cmidrule(l){3-3}
Technology (nm)               & 65                & 65                                    \\
Supply (V)                    & 1.1               & 1.2                                   \\
Max. Frequency (GHz)          & 1               & N/A                                   \\
Area ($\mu\text{m}^2$)        & \numprint{225964} & \numprint{3264}                       \\
W.C. Latency (ns)             & 2      & 3                \\
Avg. Latency (ns)             & 1       & 1.1                    \\
W.C. T/P (Gbps)               & 32    & 21.3                  \\
Avg. T/P (Gbps)               & 64    & 58.2                  \\
Code compatible               & Yes     & No                  \\
 Rate compatible              & Yes     & No                 \\
\bottomrule
\end{tabular}
\end{adjustbox}
\end{table}

\section{Conclusion}
In this paper, we propose the first hardware architecture for the GRAND-MO algorithm. GRAND-MO is a GRAND variant that is used to decode linear block codes on communication channels with memory. Since GRAND is code-agnostic, the proposed GRAND-MO architecture will decode any error correcting code provided the length and rate constraints. We suggest modifications in the GRAND-MO algorithm to simplify the hardware implementation and reduce the complexity of the decoding process. The results of ASIC synthesis show that with a code length of $128$ and a target FER of $10^{-5}$, an average information throughput of 52 Gbps and a $3$ dB gain in decoding performance can be achieved when compared to the GRANDAB (AB=3) decoder. Moreover, compared with the BCH decoder tailored for a (79,64) code, the proposed VLSI implementation achieves $33\%$ higher worst-case throughput while also providing a $2$ dB gain in decoding performance for a target FER of $10^{-5}$. In addition to that, the average  throughput for the same parameters can reach up to 64 Gbps. This proposed architecture paves the way for future soft-input GRAND-MO implementations.
\balance
\bibliographystyle{IEEEtran}
\bibliography{IEEEabrv, GRANDMO}
\end{document}